\documentclass[12pt]{article}

\usepackage[letterpaper,margin=1in]{geometry}
\usepackage[T1]{fontenc}
\usepackage[utf8]{inputenc}
\usepackage{booktabs}
\usepackage{array}
\usepackage{url}
\usepackage[hidelinks]{hyperref}

\title{CPU-Auth: Device Fingerprinting for Authentication\\via DVFS Side-Channel}
\author{Ryan Swift$^{1}$\\[0.5ex]\small $^{1}$University of California, Davis}
\date{June 6, 2024}

\begin{document}
\maketitle

\begin{abstract}
Lack of effective authentication has resulted in numerous security and privacy breaches, including unauthorized access to protected information, identity theft, and fraud. One approach to mitigating such attacks is Multi-Factor Authentication (MFA), in which users must provide multiple pieces of information for authentication. Some secondary authentication factors include SMS text verification codes, biometrics, and tokens. Each contains at least one notable flaw: SMS is notoriously insecure; biometrics rely upon access to sensitive personal data; and tokens require dependence on third-party providers (e.g. OAuth providers). This work explores CPU-Auth, a novel authentication mechanism based on unique variations in the physical characteristics of the CPU of a computing device. By measuring the behavior of the Dynamic Voltage and Frequency Scaling (DVFS) governor remotely from within a browser, unique properties of the CPU can be leveraged to establish a hardware-based device fingerprint for use in CPU-Auth. The performance of CPU-Auth is evaluated on over 50,000 data traces using distance-based and deep learning methods. CPU-Auth is part of a larger research project, and the results provided in this report reflect only the contributions made to the project by members of this group.
\end{abstract}

\section{Introduction}
In the domain of Computer Security and Privacy, the delineation between attack and defense is often obfuscated. In many instances, the same technology deployed in a system designed to provide protection can be used in another system to implement malicious attacks or adversarial behavior. The practice of fingerprinting is an excellent example of such a technology. In fact, the project from which CPU-Auth originated was conceived as a proposal for an attack which could stealthily track and identify website visitors. Given that the authentication problem is easier (binary classification, rather than multi-class classification), it was decided to first pursue an authentication implementation.

Rather than being a standalone authentication method, CPU-Auth is intended to be used as an additional factor within a MFA (Multi-Factor Authentication) scheme. It has become abundantly clear that traditional authentication systems---which typically only make use of passwords as the sole source of authentication---are inadequate for today's levels of attack frequency and sophistication~\cite{ba2018abc}. MFA has become essential for authentication in the modern world, and especially for systems which operate on sensitive, personally-identifiable information such as in the finance and medical sectors.

CPU-Auth positions itself as a tool which can be used to capture minute details and discrepancies in hardware from within the sandbox environment of a web browser. It is a system which can quickly verify identity without having physical access to the device, and without requiring elevated (i.e. sudo) permissions. CPU-Auth's device fingerprint should not be confused with a browser fingerprint, in which a fingerprint is established by collecting data about a user's browser~\cite{laor2022drawnapart}. While CPU-Auth operates within the browser, the phenomenon it captures is a distinctly hardware-based (i.e. device-based) fingerprint.

\section{Background}
\subsection{Fingerprinting}
In the context of computer security and privacy, the term \emph{fingerprinting} refers to the establishment of a signature for a given entity which makes that entity uniquely identifiable~\cite{zhang2019sensorid}. Similar to how humans can be identified by the physical fingerprint on their fingers, entities in the digital space can also be uniquely identified using various methods~\cite{zhang2019sensorid,dipta2022dfsca,laor2022drawnapart,sanchez2018clock,cheng2019demicpu}.

There are generally two important characteristics in a fingerprint: uniqueness and stability~\cite{sanchez2018clock}. Uniqueness is a measure of how likely it is that two different entities will be assigned the same fingerprint (i.e. probability of a collision). Stability is a measure of robustness, and the resistance to both passive change and active efforts which may potentially impact the fingerprint's root phenomenon. Many factors may lead to changes in a fingerprint, from the natural passage of time, to adversarial countermeasures. The goal is to have a fingerprint which is unique for every entity, and which does not ever change.

Fingerprinting methods can broadly be categorized as either software-based or hardware-based. The basis for the fingerprint refers to the root phenomenon being captured in the fingerprint. For example, the HTML5 Canvas API and WebGL API both capture information related to the browser to develop a software-based fingerprint. Alternatively, Sanchez-Rola, Santos, and Balzarotti developed a hardware fingerprint by capturing the imperfections in the quartz crystals that maintain the machine clock~\cite{sanchez2018clock}. Generally, software-based fingerprints are flexible and simpler to implement, while hardware-based fingerprints tend to be more robust and stable.

\subsubsection{Software Fingerprinting}
Software fingerprinting methodologies, such as those which make use of the Canvas and WebGL APIs, capture information about the state, structure, or usage of software~\cite{sanchez2018clock,laor2022drawnapart}. Though software fingerprints are actively in use in the wild (often deployed by websites for advertising purposes), they carry significant drawbacks which detract from their fitness for use in authentication~\cite{laor2022drawnapart}. A primary concern is the fragility and lack of robustness of software fingerprints. In our own testing, it was found that the software fingerprint registered by FingerprintJS changed after simply installing another font. Furthermore, software fingerprints which aim to increase robustness via identification of usage patterns and tracking user behavior represent a significant violation of user privacy~\cite{giura2014adaptive}. While software fingerprints are often simpler to develop and are easier to deploy remotely, their lack of robustness presents a large concern which must be addressed prior to inclusion in any authentication scheme.

\subsubsection{Hardware Fingerprinting}
Hardware fingerprints are constructed by capturing some physical property of a device which contains enough entropy to be uniquely identifiable in the target environment, and is also stable over time and in the presence of external variations and perturbations~\cite{ren2019hardware}. Relative to software-based fingerprinting techniques, the hardware basis is far more robust against relay attacks and device cloning~\cite{ren2019hardware}. Though hardware fingerprints are inherently less secure than using a dedicated hardware component to securely host applications and store cryptographic data, they are a far more scalable and cost effective solution~\cite{ren2019hardware}.

The hardware components used to construct computing devices are made to a certain manufacturing specification, within a tolerance. Therefore, any method which attempts to capture uniqueness in the physical characteristics of hardware must be sufficiently sensitive to detect such minor variations in the manufacturing process~\cite{ren2019hardware}. Often, the source of information for a hardware fingerprint lies at a modal intersection, such as when one type of signal is transformed into another, or at the interface between hardware and software~\cite{ren2019hardware}.

\subsection{Dynamic Voltage and Frequency Scaling (DVFS)}
Modern CPUs are equipped with the ability to operate at different voltage configurations and clock frequencies, depending on the system processing load~\cite{dipta2022dfsca}. Each of these voltage/frequency configurations can be referred to as performance states of the CPU, and are also referred to as Operating Performance Points (Linux kernel) and P-states (Intel CPUs)~\cite{dipta2022dfsca,wang2022hertzbleed}. Switching between performance states allows the CPU to optimize its power consumption, thermals, and performance. In performance states with a higher clock frequency, instructions can be completed faster, and more power is drawn. As power draw increases, performance increases, but the CPU becomes increasingly hot. To prevent the CPU from overheating, it must enter a lower voltage/frequency state, decreasing power draw and clock speed to maintain a safe operating temperature at the cost of performance~\cite{dipta2022dfsca}. To optimize all of the involved variables, the CPU defaults to a moderate performance state, scales up to a high performance state as needed, and scales up and down under load, as necessary to maintain safe thermal boundaries.

The mechanism which adjusts the performance states of the CPU to perform the necessary optimizations is referred to as Dynamic Voltage and Frequency Scaling (DVFS). The part of the Linux operating system which is responsible for DVFS is the CPUFreq subsystem. The CPUFreq system supports a number of scaling governors, which are the algorithms used to perform the frequency scaling~\cite{dipta2022dfsca}. These include Performance (highest frequency), Powersave (reduced frequency), and Ondemand (scale according to CPU load)~\cite{dipta2022dfsca}.

\subsection{Authentication}
Authentication typically involves two distinct phases: a registration phase and an authentication phase~\cite{ba2018abc,ren2019hardware}. In the registration phase, the user sets up the token which will be used to perform the verification. This may include setting a password, connecting an account to a rolling passcode service such as Google Authenticator, or permitting the collection of data necessary to establish a fingerprint. The token created during the registration phase acts as the baseline against which verification attempts will be compared during the authentication phase.

Later, when the user requests to perform a privileged task, such as logging in to an email account or making an online purchase, the identity of the user must be verified. Thus, the user enters the authentication phase. To be allowed to complete the privileged task, the user must supply the same (or similar, for a fingerprint) token which was set up during the registration phase. If the user fails to provide the token, or provides an incorrect one, the authentication fails and the user will be unable to complete the requested task. If the user provides the correct token, the authentication succeeds and they receive elevated privileges to complete the task.

\section{Related Work}
CPU-Auth presents a unique approach to device fingerprinting, and is informed by a large body of existing work. Some of the notable work which influenced CPU-Auth is discussed in this section, including which aspects of the selected works informed the development of CPU-Auth, and also how CPU-Auth differentiates itself.

\subsection{DF-SCA}
Dipta and Gulmezoglu~\cite{dipta2022dfsca} demonstrated with DF-SCA that DVFS could be used to perform fingerprinting. However, their goal was instead to fingerprint websites and perform an attack on user privacy by detecting which websites a user visits based on the DVFS fingerprint captured while rendering the website. While the underlying phenomenon being exploited is similar to that of CPU-Auth, the implementation and end goals are far different from DF-SCA. There are two ways in which CPU-Auth differentiates itself in its approach to DVFS fingerprinting. First, DF-SCA assumes the ability to install a malicious application on the victim's computer to collect and transmit values collected from the Linux CPUFreq interface. Access to CPUFreq is privileged (e.g. requires sudo permissions) and thus DF-SCA requires a user to install a native application and grant it elevated permissions for the attack to succeed. CPU-Auth works in-browser, without requiring elevated permissions or the installation of a separate application, yet captures the same phenomenon as DF-SCA. Second, CPU-Auth fingerprints devices and classifies on the basis of similarity to a reference, whereas DF-SCA classifies based on the reference itself. One of the outcomes of this decision is that, while CPU-Auth only needs to learn binary classification for similarity, DF-SCA has to train a multi-class model which can account for a wide range of combinations of microarchitectures, DVFS governors, URLs, and browsers~\cite{dipta2022dfsca}. CPU-Auth is a much simpler approach than DF-SCA, requiring no elevated permissions, and performing simpler binary classification, yet captures the same underlying behavior of DVFS as DF-SCA.

\subsection{DeMiCPU}
Similar to CPU-Auth, DeMiCPU~\cite{cheng2019demicpu} also performs device fingerprinting and aims to capture the physical discrepancies in CPUs leftover from the manufacturing process via a side-channel. The primary manner in which CPU-Auth differentiates itself from DeMiCPU is that, rather than collecting timing values as a proxy for DVFS behavior, DeMiCPU directly reads the magnetic induction (MI) portion of the electromagnetic interference (EMI) emitted by the device. The consequence of this approach is that DeMiCPU fingerprinting cannot be performed without physical access to the device being fingerprinted and a special piece of hardware, which presents limitations on its practical utility and capacity for wide deployment. Additionally, DeMiCPU presents results only from experiments performed in a controlled laboratory setting, and with a maximum of 90 devices~\cite{cheng2019demicpu}. CPU-Auth, on the other hand, can be deployed remotely without access or hardware limitations, and presents results based on an in-the-wild deployment involving over 50,000 devices.

\subsection{DrawnApart}
Of the three papers discussed in this section, DrawnApart~\cite{laor2022drawnapart} is the most similar approach to CPU-Auth. DrawnApart positions itself as a technique which can increase the length of time for which browser fingerprint-based tracking is effective~\cite{laor2022drawnapart}. Similar to CPU-Auth, DrawnApart is a stateless approach, and does not require setting cookies or storing any other information on a user's computer. Though DrawnApart claims to capture a browser fingerprint, the underlying phenomenon being captured to produce the fingerprint is actually based on the behavior of the graphics processing unit (GPU)---a hardware component. Additionally, rather than presenting itself as a standalone fingerprinting technique, DrawnApart is instead intended to be used as part of a larger fingerprinting scheme, and reports its results as the lift or marginal gains which it provides when used in combination with a selection of popular existing browser fingerprinting methods~\cite{laor2022drawnapart}. At a high level, CPU-Auth is quite similar to DrawApart: it is a fingerprinting method which captures unique variations in hardware characteristics---as observed in timing values---to establish a fingerprint remotely from within a browser. CPU-Auth differentiates itself by fingerprinting the CPU and DVFS behavior, and positions itself as a device fingerprinting technique, presenting its results in classification performance, rather than lift.

\section{System Model}
For illustrative purposes, the example deployment of CPU-Auth will be in the fraud prevention system of a bank. For illustrative purposes, let Alice be a user of such a bank. Upon opening an account with the bank, Alice registers her phone with the bank, allowing the bank to store her phone's fingerprint. When a purchase is made using Alice's bank account which may appear to be risky or unusual---such as an online purchase made from an IP address which Alice doesn't usually use---Alice's bank will challenge the user to provide proof of identity. If Alice is indeed the one who attempted the transaction, when she accepts the bank's challenge to run an authentication check on her verified phone, CPU-Auth will detect the high similarity between the registered fingerprint and query fingerprint and approve the transaction. Should an unauthorized user attempt to make a purchase on Alice's account, CPU-Auth will detect that the query fingerprint is excessively different from Alice's registered phone fingerprint, and deny the transaction.

\section{Threat Model}
Once again, let Alice be a user of a bank which makes use of CPU-Auth to detect and prevent fraud. Also, let Mallory be a malicious adversary who aims to make a fraudulent purchase using Alice's bank account. In the CPU-Auth threat model, there are two assumptions made. First, it is assumed that Mallory is in possession of Alice's credit card information, along with any additional information necessary to initiate a purchase using Alice's account. Second, it is assumed that Mallory is not present during the time at which Alice registers her device with the bank, and that the fingerprint which the bank will use to establish the reference for Alice's verification is indeed correct and valid.

\section{Methodology}
\subsection{Power Viruses}
In order for CPU-Auth to capture the DVFS information necessary to establish a fingerprint, it needs to observe multiple changes in DVFS state. To facilitate this behavior, CPU-Auth employs a power virus to force CPU frequency scaling. The power viruses developed for CPU-Auth use a crafted, intense simulated workload (repeated floating-point operations) to push the CPU toward a performance state, and periods of sleep to allow the CPU to drop its frequency state. The data used in this analysis were generated using CPU-Auth's step virus, which alternates between executing a constant workload and sleeping. The step virus was designed to induce a high number of switches in performance states, as one of the key factors CPU-Auth seeks to capture is DVFS behavior during state switches.

\subsection{Timing Mechanism}
CPU-Auth's goal of fingerprinting the behavior of DVFS within the sandboxed environment of a browser presents a significant challenge~\cite{greamo2011sandboxing}. In particular, measuring the relationship between the DVFS governor and power and frequency data is difficult when CPU-Auth has access to neither the particular DVFS state, nor the power and frequency data. As a result, CPU-Auth is instead designed to measure elapsed time of computation, as this is directly related to the frequency and power usage of the CPU, and is measurable within a browser at millisecond resolution~\cite{irishPerformance,mdnPerformance}.

The timing mechanism in CPU-Auth, henceforth referred to as the clock, works by executing a standardized, fixed computation and recording the elapsed time between the start and end of that computation. The standard computation used for the clock was 1000 executions of a random number generator (RNG), which resulted in roughly ten-millisecond scale results in preliminary testing on thin desktop clients. The clock runs concurrently with the power virus, collecting timing values into an array, as the power virus induces state switches in DVFS. As the power virus pushes the governor into a performance-oriented state, clock values decrease. This is due to the CPU boosting its power draw and operating frequency, resulting in faster (i.e. lower) computation times. Similarly, in the virus's sleep periods, or when the CPU is forced to throttle due to thermal limitations, the DVFS governor drops the power draw and CPU frequency to a lower-performance state, preventing the CPU from executing the clock computations as quickly, and producing slower (i.e. larger) clock values.

One of the primary implementational challenges of CPU-Auth was ensuring that the power virus and clock are executed on the same thread. The JavaScript virtual machine does not expose direct relation between Web Workers (the JavaScript API used for multithreading) in the browser and the utilization of logical cores in the processor~\cite{verdu2016webworkers}. Because of this, there is little control over the physical location of thread execution from within the browser. One of the side-effects of this is that the power virus and clock may be scheduled on different physical threads. Given that CPU-Auth is trying to acquire a representation of the physical characteristics of a CPU, it is important that the clock and virus be scheduled with consistent physical proximity on the CPU. The admittedly inelegant solution employed by CPU-Auth is to schedule as many Web Workers for the power virus as there are physical threads on a given device's CPU. Thus, regardless of where the clock is scheduled, it will always be on the same thread as an instance of the virus, and all other threads will also be executing an instance of the virus.

\subsection{Data Collection}
The data used in the experiments conducted for this paper was collected from an in-the-wild deployment of CPU-Auth's fingerprinting methodology. A website was created such that, upon visiting the site, the fingerprinting process would begin automatically. During this process, the power virus runs, timing values are collected, and a selection of other information about the device is captured, including HTTP headers and the FingerprintJS output. Additionally, a cookie is set in the visitor's browser to serve as ground truth during analysis. The cookie can be understood as the device identifier or ID, and thus, should two time series traces be collected with the same cookie, it is concluded that they originated from the same device. Due to the presence of sensitive information in the collected data, hashed values of the cookie, visitorId, and IP address were stored, so as to protect user privacy. In total, over 50,000 data traces were collected. Due to the nature of the authentication problem, at least two traces from a given device are necessary to form a positive pair. 225 different devices were found to have produced at least two traces, with as many as 120 traces from a single device.

\subsection{Performing Authentication}
The authentication problem in CPU-Auth is formulated as a binary classification task: Given two fingerprints, are they from the same device or different devices? When one of the two fingerprints is fixed as a reference, and the other is a query fingerprint---meaning the origin of the fingerprint is uncertain---the authentication problem can be restated as such: Does the query fingerprint come from the same device as the reference, or not? To perform such a classification, a metric must be used to determine similarity between two fingerprints. Also, since it is exceedingly unlikely that any device would produce the exact same fingerprint (e.g. time series), it is necessary to compute the similarity between two fingerprints and compare that value against a threshold. For example, Euclidean distance is often used to quantify similarity, and thus any computed distance lower than a fixed threshold would be considered sufficiently similar and thus from the same device, and distance values greater than the threshold would be considered to have originated from two different devices.

Unfortunately, Euclidean distance is unsuited for CPU-Auth, as the data collection process resulted in variable length time series. Though these time series could be padded or truncated to perform the distance computation, preliminary experiments showed extremely poor performance. Instead, four other approaches to classification were used to determine which performs best for CPU-Auth. These approaches were adapted from the methods developed by Wang, Yan, and Oates~\cite{wang2017timeseries} to work in the similarity-based class definitions of CPU-Auth. Of the four methods tested, one was a simpler, non-ML method, and the other three employed simple deep learning architectures. The distance-based metric Dynamic Time Warping (DTW) was used, as it has the ability to handle variable and unequal length time series. DTW is generally understood to be a standard metric used for time series classification, as it is simple, yet often highly performant~\cite{fawaz2019deep}. In a departure from the work done in~\cite{wang2017timeseries}, CPU-Auth's DTW implementation sets a threshold for similarity values, rather than performing a 1-nearest-neighbor (1NN) classification.

Additionally, CPU-Auth's problem formulation necessitates a modification to the deep learning implementations. Since there are no representative fingerprints for the two classes (same device and different devices), the deep learning architectures must instead be constructed to perform some form of similarity learning. The top level of the networks learn to extract a lower dimensional embedding, which is then fed through a binary classifier multi-layer perceptron (MLP), and trained using binary cross-entropy loss with PyTorch's AdamW optimizer. In training, the top-level embedding extractors learn to produce a representation of the traces which allows the binary classifier to make the distinction between traces which originate from the same and from different devices.

Data preprocessing involved constructing pairs of time series traces, where each pair was assigned a label of 1 for a positive pair and 0 for a negative pair. ``Positive'' indicates that the traces originated from the same device and ``negative'' indicates that the traces came from different devices. These pairs were formed by creating positive and negative lists of traces for each device by determining the number of traces collected for a given device, selecting a random sample of traces from any other device in the dataset that is equal in number to the traces from that device, and using Cartesian product to form lists of positive and negative pairs. The product of the positive list with itself forms the positive pairs, and the product of the positive list with the negative list forms the negative pairs. Thus, there are an equal number of positive and negative pairs for every device. For DTW, the traces in these pairs were formed using the raw time series data. For the deep models, the traces were first padded or truncated to a fixed length of 2000, then scaled with Z-score normalization. Additionally, the training and validation splits were created by partitioning 80/20 by device ID. Therefore, any device present in the training set is absent in the validation set, and vice versa.

The three base deep models used were also adapted from~\cite{wang2017timeseries}. CPU-Auth uses a four-layer MLP, a fully convolutional network (FCN), and a three block ResNet which uses a modified (i.e. much smaller) FCN as the core of each block. Each of these models outputs a linear embedding of size 128. To perform the similarity comparison, each of these models is run in a Siamese network, wherein each time series in a pair passes through the model separately. The two resulting embeddings are concatenated, then passed through a three-layer binary classifier. Additional details of the architectures developed for CPU-Auth are listed in Appendix~A.

\section{Results}
\begin{table}[htbp]
\centering
\begin{tabular}{lrrrrr}
\toprule
Base Model & \begin{tabular}[c]{@{}c@{}}Model Parameters\\(incl. classifier)\end{tabular} & Accuracy & F1 Score & Precision & Recall \\
\midrule
DTW    & N/A       & 0.6914 & 0.6918 & 0.6909 & 0.6926 \\
MLP    & 2,181,505 & 0.9500 & 0.9474 & 0.9000 & 1.0000 \\
FCN    & 33,090,433& 0.9250 & 0.9455 & 0.8966 & 1.0000 \\
ResNet & 8,255,297 & 0.5250 & 0.6885 & 0.5385 & 0.9545 \\
\bottomrule
\end{tabular}
\caption{Results of CPU-Auth using different base models.}
\label{tab:results}
\end{table}

The results indicate that the MLP is the most successful classifier in every category, with the FCN being not too far behind. Although the FCN had a slightly lower accuracy than MLP, the FCN nearly perfectly matched the MLP's F1 score, precision, and recall. DTW produced the smallest difference between precision and recall, while the ResNet possessed the largest difference. The ResNet was the worst performer in every category except recall. Notably, the ResNet's accuracy was approximately 17 percentage points lower than the DTW baseline.

\section{Discussion}
As a point of reference, the accuracy of binary classification at random is 0.5000. Higher precision values indicate that the given method is effective at classifying positives, resulting in minimal false positives. Higher recall indicates effectiveness in classification of negatives. The F1 score is the harmonic mean of precision and recall, combining both metrics into a single value. For each of the three deep models, the recall is significantly higher than the precision, which implies that future work may be able to extract even better performance from the models if they learn to sacrifice a bit of performance in negative classification for increased precision. Although the MLP performed the best on the validation set, the nearly identical values of the FCN indicate that it may generalize similarly well to the MLP, and thus should not be written off as inferior to the MLP.

One undeniable difference between DTW and the deep models is the complexity of implementation and training. DTW is simply distance with a threshold, whereas each model has to learn millions of parameters. Though the MLP and FCN were able to take advantage of their depth to learn some of the nonlinearities missed by DTW, the ResNet had a difficult time performing better than a random guess. With each of the three blocks in the ResNet possessing the same depth as the entire embedding extractor of the FCN, the ResNet may simply have been too deep of a model, as was the case in~\cite{wang2017timeseries}. Despite the far smaller parameter count relative to the FCN, generalization was never achieved with the ResNet. Each of the three models struggled significantly with overfitting, and relatively large weight decay values were needed for regularization. The ResNet was especially tricky to tune, and only a marginal improvement over random was able to be extracted from the model. Though certainly beaten by the MLP and FCN, DTW continues to demonstrate that its simple approach can be useful and effective in the task of time series classification.

\section{Conclusion}
CPU-Auth is a hardware fingerprinting method which can be deployed remotely within the browser and has demonstrated up to 95\% accuracy discriminating between fingerprints extracted from the same and from different devices. The potential for successful generalization for the MLP and FCN models is likely quite high, as the models were both able to achieve over 92\% accuracy without any information about the particular type of device being tested. DTW is found to be a respectable baseline, providing approximately a 20 percentage point lift over random, and significantly outperforming the ResNet. The very deep ResNet model had trouble learning generalizable information from the training data, and achieved only 52\% accuracy on the validation set. Implementations of CPU-Auth with either the MLP or FCN have been demonstrated as over 92\% accurate in device-based authentication.

\section{Future Work}
Three tasks are left to future work. First, making the distinction between two devices is easiest at the highest level of granularity, where all devices are equally likely to be compared against each other. A harder problem is discriminating between devices of the exact same type---two iPhone 15's, for example. Future work should ensure that accuracy remains sufficiently high as devices being compared become more and more similar. Second, CPU-Auth is susceptible to replay attacks. Should a fingerprint transmission be intercepted, an adversary would be able to impersonate a user by sending the intercepted fingerprint, rather than that of the adversary's own device. Future work should construct a more robust system model which is resistant to replay attacks. Finally, the stability of CPU-Auth's fingerprint has yet to be sufficiently evaluated. Understanding the effects of factors such as time since registration, ambient temperature, ambient humidity, background computational load, and use of varying browsers may contribute to a more robust authentication pipeline.

\appendix
\section{Appendix: Implementation Details}
\subsection{Data Splits}
\begin{table}[htbp]
\centering
\begin{tabular}{lrr}
\toprule
& Train & Validation \\
\midrule
Number of Pairs in Split & 88,072 & 9,448 \\
\bottomrule
\end{tabular}
\caption{Number of total pairs in Train and Validation splits. The 80/20 split on devices between Train and Validation resulted in approximately a 90/10 split in pairs.}
\end{table}

\subsection{DTW Threshold}
\begin{table}[htbp]
\centering
\begin{tabular}{lr}
\toprule
DTW Threshold & 71200.0 \\
\bottomrule
\end{tabular}
\caption{The threshold learned from the Train split and used to achieve the results in Validation for classification based on DTW.}
\end{table}

\subsection{Deep Model Hyperparameters}
\begin{table}[htbp]
\centering
\begin{tabular}{lrrr}
\toprule
& MLP & FCN & ResNet \\
\midrule
Learning rate (base model) & 1e-5 & 2e-7 & 1e-6 \\
Learning rate (classifier) & 1e-4 & 1e-6 & 1e-6 \\
Weight decay (base model) & 1e-1 & 1e-2 & 1e-3 \\
Weight decay (classifier) & 1e-1 & 1e-2 & 1e-4 \\
Batch size & 64 & 64 & 64 \\
Epochs & 5 & 4 & 5 \\
\bottomrule
\end{tabular}
\caption{Hyperparameters used to achieve final results for each combination of base model and classifier in the three CPU-Auth implementations.}
\end{table}

\subsection{Deep Model Architecture Details}
\subsubsection{Multilayer Perceptron}
The MLP embedding extractor is a 4-layer MLP which takes input traces of length 2000. The output dimensions of the 4 layers are 512, 1024, 512, 128. There is dropout before each of the 4 layers, with values 0.4, 0.3, 0.4, 0.3. After each layer, Leaky ReLU is applied.

\subsubsection{Fully Convolutional Network}
The FCN embedding extractor has 3 convolutional layers and a final linear layer. The number of input channels is 1 since CPU-Auth analyzes univariate time series, and the output channels in the 3 convolutional layers are 128, 256, 128, with kernel sizes 7, 5, 3 and padding values of 3, 2, 1, and the rest of the parameters are set to the PyTorch defaults. After each convolutional layer is a 1D batch norm and a Leaky ReLU. The output of the final convolutional layer is flattened and sent through a linear layer wrapped in another Leaky ReLU, reducing the output dimension to 128.

\subsubsection{ResNet}
The ResNet embedding extractor is made up of 3 blocks, each with 3 convolutional layers, plus a separate convolutional layer to compute the residual for each block. The same kernel and padding values are used from the FCN. The first convolutional layer in each block changes the number of channels between the input and output, and the number of output channels is consistent for the rest of the block. With 1 input channel, the output channels in each block are 16, 32, 32. The convolutional layer to perform the transformation of the input for the residual calculation is similar to the first layer in each block, but with a kernel size of 1 and default padding. After passing through each block, a batch norm is applied before a Leaky ReLU is wrapped around the residual. At the end of the ResNet, the output is flattened and passed through a linear layer wrapped in a Leaky ReLU, reducing the dimension to 128.

\subsubsection{Binary Classifier}
The binary classifier is a 3-layer perceptron, and the exact same classifier is placed at the end of each of the embedding extractors. Its inputs are the 2 embeddings (one from each trace in the pair) and it outputs a value between 0 and 1. The input size is 256 (128 for each embedding) and the output sizes of each layer are 128, 64, 1. Dropout is used before each layer, with rates 0.3, 0.4, 0.2. Leaky ReLU is applied at each layer, and a Sigmoid activation is applied to the output of the final layer to produce a result which is between 0 and 1. This result is rounded to perform classification and analysis.

\bibliographystyle{plain}
\bibliography{references}

\end{document}